\documentclass[preprint,showpacs,preprintnumbers,amsmath,amssymb]{revtex4}
\usepackage{graphicx}
\usepackage{dcolumn}
\usepackage{bm}

\begin{document}


\title{Acceleration by oblique shocks at supernova remnants
\\and cosmic ray spectra around the knee region}

\author{K. Kobayakawa}
\email{kobayakw@ccmails.fukui-ut.ac.jp}
\affiliation{%
Fukui University of Technology, 
Gakuen 3-6-1, Fukui, Fukui, 910-8505, Japan
}%

\author{Y. S. Honda}%
\email{yasuko@ktc.ac.jp}
\affiliation{%
Department of Electrical and Information Engineering, 
Kinki University Technical College, 
2800, Arima, Kumano, Mie, 519-4395, Japan
}%

\author{T. Samura}
\affiliation{
Department of Electrical and Computer Engineering, 
Akashi National College of Technology,\\
679-3, Nishioka, Uozumi-cho, Akashi, 674-8501, Japan
}%

\date{\today}

\begin{abstract}
We examine the first order Fermi acceleration on the presumption 
that supernova remnant shocks cross ambient magnetic fields with 
various angles. These oblique shocks accelerate particles more 
efficiently than the parallel shocks and elevate the maximum 
energies achievable by the particles. 
The primary cosmic ray spectrum is strongly dependent upon these
energies. 
We also consider the dependence of the injection efficiency 
and of spectral indices on obliquity. 
When indices and absolute fluxes at $10^{12}$ 
eV are given for six nuclear groups from balloon-borne 
data, each energy spectrum develops a smooth rigidity dependent
knee structure. The resultant total 
spectrum also behaves similarly and fits well with 
ground-based experimental data up to several  
$10^{17}$ eV.
It is shown as well that the chemical composition changes 
significantly from lighter to heavier nuclei as the energies 
of particles exceed the knee region. Other predicted curves are 
compared with the experimental data which they reproduce rather well.
\end{abstract}

\pacs{PACS 96. 40. De, 96. 50. Fm, 98. 70. Sa. }
\maketitle

\section{\label{sec:level1} Introduction}
In recent years, many measurements of the primary
spectrum and chemical composition of cosmic rays have 
been carried out. Some of them are performed directly by 
balloon-borne experiments up to about 100 TeV. Others are by 
indirect ground-based experiments ranging from sub-TeV 
to some 100 EeV. These experiments have mainly
been done to study the knee region. 
Recent reports indicate that the all-particle spectrum of cosmic 
rays does not present a sudden steepening, but rather a gradual bent near 
the energy at the knee, $E_{{\rm knee}}=(2-5)\times 10^{15}$ 
eV, from $E^{-(2.6\sim 2.8)}$ to $E^{-(3.0\sim 3.2)}$ 
\cite{amenomori}. This value of $E_{{\rm knee}}$ 
is smaller than about a factor of 2 compared to those of 
earlier reports \cite{nagano}. 
The chemical composition of cosmic rays below $E_{{\rm knee}}$ is 
directly measured and is almost constant in terms of the
average logarithmic mass, $<\ln A>$. However, the data of 
$<\ln A>$ above $E_{{\rm knee}}$ fairly scatter, because they 
are obtained by indirect air shower observations. It is also 
noted that the value of 
$<\ln A>$ varies depending on the hadronic interaction 
models assumed in simulations of extensive air showers 
\cite{casablanca}.

On the other hand, it is generally believed that cosmic rays 
in the energy range up to $E_{{\rm knee}}$ are accelerated 
mostly by a supernova remnant (SNR) shock although 
there still exist some arguments that the origin of 
galactic cosmic rays (GCRs) is not yet settled \cite{parizot}.  
When particles cross the shock front back and forth
frequently, they obtain high energies proportional to the 
shock velocity, which is called the first order Fermi 
acceleration (for 
review, see \cite{blandford}). This stochastic acceleration and 
deformation of source spectra due to energy-dependent diffusion 
from the galaxy can explain the power-law spectrum up to the knee. 
This scenario predicts that the maximum energy $E_{{\rm max}}$, 
that particles can achieve by the shock acceleration mechanism, 
nearly equals $E_{{\rm knee}}$. 

Actually $E_{{\rm max}}$ is estimated to be $\sim 3Z\times 
10^{13}$ eV \cite{gaisser} at most, where $Z$ is the atomic 
number, which is nearly two orders lower than $E_{{\rm knee}}$. 
However, this value is 
achieved by parallel shock acceleration where the normal of the
shock front is parallel to the direction of the interstellar magnetic 
field. This is  different from nonparallel shocks, where the angle 
$\alpha_{1}$ between the shock normal and the field direction 
is not zero. 
These oblique shocks can accelerate particles much more 
efficiently \cite{jokipii,ostrowski88,tt91}.
For extremely oblique (quasiperpendicular) shocks, 
the values of $E_{{\rm max}}$ are increased by several orders of magnitude 
compared to those in parallel collisions. 
The details will be 
described in the next section.

In this paper we will explain the knee behavior of 
the energy spectrum with the acceleration mechanism 
as caused by oblique shocks. Our model is based 
on a simple idea as follows \cite{kobayakawa}. 
The ejected materials from a supernova explosion expand  
into the interstellar medium (ISM) driving a shock wave. 
While the expanding shell sweep up its own mass of the ISM, 
the shock acceleration may be active. 
The ejected mass ($\sim 10$ $M_{\odot}$) with large 
momentum moves out freely at constant velocity $10^{8}-10^{9}$ 
cm/s during the free expansion phase. The 
shape of the shock front is supposed to be almost 
spherically symmetric. The directions of the interstellar 
magnetic field lines will be, over a wide range, nearly random rather 
than well aligned. It is therefore assumed that the field lines
meet the shock front at random angles $\alpha_{1}$, the cosines
of which, $\eta$, are uniformly distributed. 
A more detailed discussion will be presented in section III. 
Since $E_{{\rm max}}$ is strongly dependent upon $\eta$,
$E_{{\rm max}}$ values extend from $E_{{\rm max}}(\eta = 1)
\equiv E_{\rm crit} $ in parallel shocks to  
$E_{{\rm max}}(\eta = \eta_{\rm min}) \equiv E_{\rm cut}$ in 
quasiperpendicular ones. The uniform distribution of $\eta$
permits that particles with $E,\,E_{\rm crit} < E < E_{\rm cut}$, 
have less chances to reach high energies 
than those with $E \leq E_{\rm crit}$ . 

There is an important question in acceleration at 
oblique shocks \cite{baring93,baring94,ellison95}, 
however, namely that the rate of accelerated 
particles (injection efficiency) decreases drastically 
as the shock obliquity increases.
On the other hand, the obliquity produces a harder energy spectrum
\cite{ostrowski91,ostrowski93,naitop,naitom}.
Since particles with $E<E_{{\rm crit}}$ can be considered to be
accelerated at shocks with the whole range of $\eta$, we do not apply
modifications to the given spectra.
For $E > E_{\rm crit}$, both corrections of the injection
efficiency and the change of the spectral index are taken into
consideration, in addition to the chance probability of having $\eta$.
Their fluxes decrease and cause a gradual
change of spectral indices. In terms of rigidity, 
$R_{\rm crit}$ and $R_{\rm cut}$, corresponding to 
$ E_{\rm crit}$ and $ E_{\rm cut}$, are the same irrespective 
of $Z$. The flux of each elemental group starts to bend 
at different $ E_{\rm crit}$. Then the total flux
shows a slow change around the knee.  
 
We first choose the spectral indices and absolute fluxes for six  
nuclear groups at 1 TeV as an assumption so as to fit balloon-borne 
data. Next, based on our model, we calculate the energy spectrum 
up to $10^{6}$ TeV for each group 
and obtain therefrom the total spectrum 
and energy dependence of the chemical composition directly. 
These results are compared with recent data of balloon and air 
shower experiments.

\section{The maximum energy}
In this section we estimate the maximum energy of particles 
accelerated by oblique shocks at supernova remnants. If 
magnetic field lines are 
not parallel to the shock normal, an electric field is 
induced in the rest frame of the shock. Particles drift along 
the shock front gaining energies due to this field. In order 
to describe the interactions of particles with shocks 
concisely, the de Hoffmann-Teller frame in which the electric 
field vanishes globally is widely used \cite{drury}. 
For oblique shocks, reflection at the shock front is more 
important than diffusion in the downstream region (Kirk \& 
Heavens 1989 \cite{kirk89}), so that a more rapid acceleration 
can occur than for parallel shocks. The maximum 
energies of accelerated particles can be even a few orders of 
magnitude larger in the case of a quasiperpendicular shock 
compared to parallel ones. 

\begin{figure}[ht]
\begin{center}
\includegraphics{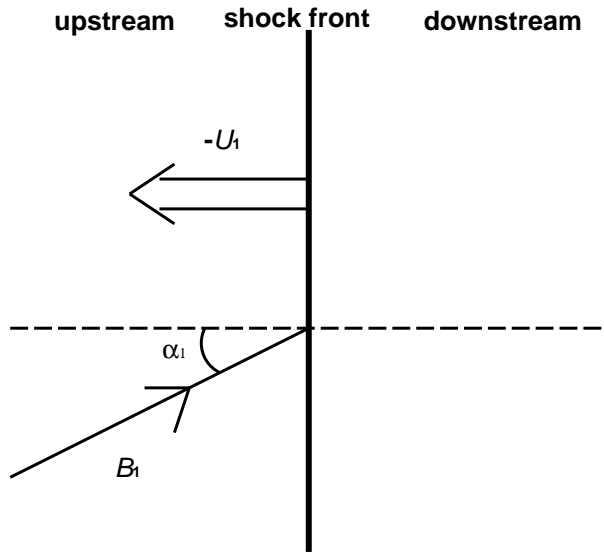}
\caption{A schematic diagram of an oblique shock 
in the upstream rest frame where the 
magnetic field $B_{1}$ intersects the 
shock normal at a nonzero angle $\alpha_{1}$.}
\label{fig:frame}
\end{center}
\end{figure}

Suppose a plane shock wave propagating at a 
non-relativistic speed $U_{1}$ with respect to the upstream 
medium. The shock normal intersects with the upstream magnetic 
field $B_{1}$ at an angle $\alpha_{1}$. A 
schematic view of the shock in the upstream rest frame is shown in 
Fig. \ref{fig:frame}. Parameters for the downstream region in its own rest 
frame, with magnetic field inclination $\beta_{2}$ and shock 
velocity $U_{2}$, can be expressed in terms of the 
upstream ones \cite{hudson}.   
We assume that the amplitude of the magnetic field perturbations 
is sufficiently small in the following calculation.

\subsection{The dependence of $E_{{\rm max}}$ on magnetic 
field inclination}
There are some energy loss processes significant for the limitation of 
particle acceleration. For example, ionization, 
bremsstrahlung, and synchrotron radiation are conceivable. 
These processes for protons or other nuclei, however, 
are not so important 
within the energy range in which we are interested. The time 
scale of acceleration by oblique shocks at SNRs is regarded as 
sufficiently small compared to those of such loss processes. 

Unless the energy losses of particles are significant, 
the finite lifetime of the supernova blast wave as a 
strong shock mainly limits the maximum energy per particle. 
Effective shock acceleration can occur 
until the swept up mass reaches the ejected mass 
$\sim 10$ $M_{\odot}$. The lifetime of the shock 
$t_{{\rm sh}}$ is estimated to be about thousand years 
for ejecta expanding at velocity $\sim 10^{8}$ 
cm/s into a medium of average density 1 proton$/{\rm cm}^{3}$. 
As long as the characteristic length 
for diffusion is much less than the radius of the 
shock, the plane approximation is valid. Balancing 
the acceleration time scale with $t_{{\rm sh}}$ after 
Lagage \& Cesarsky \cite{lagage},
we can write down the following expression for $E_{{\rm max}}$, 
\begin{equation}
E_{{\rm max}}=\frac{R_{{\rm sh}}(r-1)}{rcx}U_{1}eB_{1}Z 
\left[\cos^{2}\alpha_{1}
+\frac{\sin^{2}\alpha_{1}}{x^{2}}
+\frac{r(\cos^{2}\alpha_{1}+\frac{r^{2}}{x^{2}}\sin^{2}\alpha_{1})}
{(\cos^{2}\alpha_{1}+r^{2}\sin^{2}\alpha_{1})^{\frac{3}{2}}}\right]^
{-1},
\label{eqn:emax1}
\end{equation}
where $r$ is the compression ratio of the shock, i.e. 
$r = U_1/U_2$, and $x$ is the square root of the ratio of
$ \kappa_{\parallel}$ to $\kappa_{\perp} : x^2 = 
\kappa_{\parallel}/\kappa_{\perp}$, where the two $\kappa$'s are
the parallel and normal components of the diffusion coefficient
with respect to the magnetic field lines.

Substituting the various constants with plausible values, 
we obtain $E_{{\rm max}}$ as a function of $\eta(\equiv
\cos\alpha_1)$ and $x$,
\begin{eqnarray}
E_{{\rm max}} &=& 2.51\times10^{16}[{\rm eV}]
\left(\frac{B_{1}}{30{\rm \mu G}}
\right)\left(\frac{R_{{\rm sh}}}{3{\rm pc}}\right)
\left(\frac{U_{1}}{10^{7}{\rm m/s}}\right) \nonumber \\
& & \frac{Z}{x}\frac{r-1}{r}\left\{\eta^{2}+\frac{1-\eta^{2}}{x^{2}}
+\frac{r[\eta^{2}+\frac{r^{2}}{x^{2}}(1-\eta^{2})]}
{[\eta^{2}+r^{2}(1-\eta^{2})]^{\frac{3}{2}}}\right\}^{-1}.
\label{eqn:emax2}
\end{eqnarray}

The maximum energy for parallel shocks ($\eta=1$), 
$E_{{\rm crit}}$ can be estimated from Eq. (\ref{eqn:emax2}) 
in the case of strong shocks ($r = 4$) with $x = 30$ 
and with typical values of $B_{1}$, $R_{sh}$ and 
$U_{1}$,
\begin{equation}
E_{{\rm crit}}= 1.25Z\times 10^{14}\,[{\rm eV}].
\label{eqn:rigidity}
\end{equation}

Here we evaluate the numerical value $2.51\times 10^{16}$
according to the recent papers.  
Namely, due to non-linear effects, the maximum energies 
are enhanced by the factor of $80/9$ compared to
ordinary values \cite{berezhko96}. We will comment on 
the rather large value of $B_{1}$ \cite{bykov,klepach} 
in the next subsection.  
It is noted that the value $1.25\times 10^{14}$ 
in Eq. (\ref{eqn:rigidity})
is larger than the usual ones $ \sim 3 \times 10^{13}$. 

It is also compared to Berezhko's values \cite{berezhko96}. 
When an enhancement factor of $80/9$ is used, this value is 
larger by one order of magnitude, $\sim 1\times 10^{15}$. 
Even if replacing the parameters of his Eq. (36),
where the enhancement factor due to weak shock 
modification is $10/3$ by our typical values mentioned 
above and after the normalized value of the field 
3 mG is corrected to $3$ ${\mu}$G \cite{jetp96}, we get 
$\sim 1.4\times 10^{15}$.  
Thus in other words we may say that the value in 
Eq. (\ref{eqn:rigidity}) is only about ten times larger for field
values used in a conventional 
derivation of $E_{{\rm max}}$. Therefore numerical values in 
Eqs. (\ref{eqn:emax2}) and (\ref{eqn:rigidity}) are not 
inadequate even if $x$ has a strong effect.

\begin{figure}[ht]
\begin{center}
\rotatebox{270}{\scalebox{0.5}[0.5]{\includegraphics{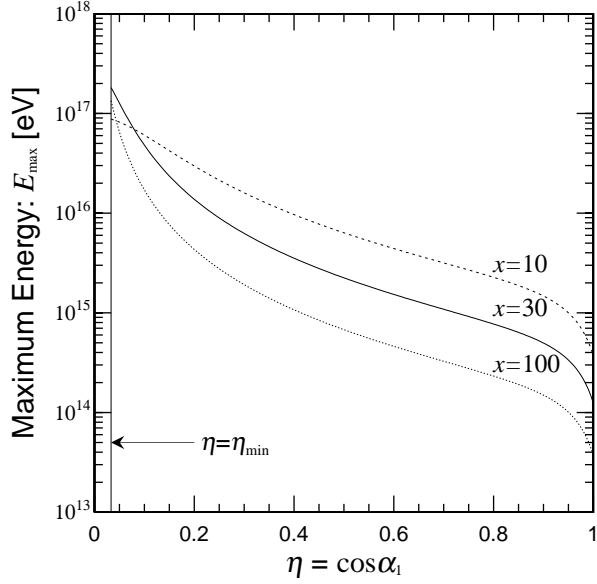}}}
\caption{The maximum energy $E_{{\rm max}}$ for a proton with 
three values of $x$ versus magnetic field inclination $\eta$. 
$\alpha_{1}$ is an angle between the magnetic field and the 
normal of the shock front. The $E_{{\rm max}}$ at $\eta=1$ 
and $\eta=\eta_{{\rm min}}$, are the energies obtained by 
parallel shock and quasiperpendicular shock acceleration, 
respectively. For the nucleus with the atomic number $Z$, 
$E_{{\rm max}}$ should be multiplied by $Z$.}
\label{fig:emax}
\end{center}
\end{figure}

In order to see how $E_{{\rm max}}(\eta)$ varies with different 
$x$ values, $x = 10, 30, 100 $ are tested since the value of $x$ 
is considered to lie between 10 and 100. Other parameters are
the same as the above ones.  Figure \ref{fig:emax} shows the $\eta$ dependence 
of $E_{{\rm max}}$ for $Z=1$: proton. 
From Eq. (\ref{eqn:emax2}), the value of $E_{{\rm max}}$ 
is proportional to the particle charge $Z$. 
As shown in Fig. \ref{fig:emax}, the $E_{{\rm max}}$ value is 
strongly dependent on the field inclination $\eta$. $E_{{\rm max}}$ in 
quasiperpendicular shocks is larger than that in parallel shocks 
by two or three orders of magnitude for each value of $x$. 
In the case of $x = 30,\, E_{\rm cut} = 1.84\times10 ^{17}$ eV 
is about 1470 times the value $E_{\rm crit}$ 
with $Z = 1$ in Eq. (\ref{eqn:rigidity}).

\subsection{Magnetic field in the supernova remnants}
As we have mentioned above, we take a $B_{1}$-value 
somewhat larger than the average strength of the 
interstellar magnetic field of magnitude of microgauss. 
It is widely known that SNRs 
are usually accompanied with magnetic fields. 
The sufficiently strong field is possibly present 
in the accelerated region around SNR-shocks 
because of rapid expansion of ejecta into the interstellar medium. 
There is some possibility \cite{lucek00} 
that cosmic ray streams amplify non-linearly magnetic fields
up to 1 mG.
Recent measurement also suggest that some galactic SNRs possess 
intense magnetic fields of the order from milligauss 
to sub-milligauss in the downstream region
\cite{yusef,claussen,koralesky,brogan}. 
These observed values are suppressed by a factor of 
5 or so in the upstream, and hence our choice 
$B_{1}=30$ $\mu$G is quite relevant.  

There are some investigations related to the magnetic field 
geometry in SNRs. In young remnants, the ambient field lines 
are stretched and amplified to be predominantly parallel to 
the shock normal on account of Rayleigh-Taylor instabilities 
\cite{jun}. On the other hand, for older SNRs 
perpendicular shocks likely originate from the compression 
of the ambient interstellar field. 
Observations of synchrotron radiation also support 
this morphology \cite{dickel,miline}. The value
of $B_1$ of our choice and isotropic distribution 
of $\eta$ may be appropriate
as far as we are concerned with the whole behavior 
of cosmic rays.

Here we also comment on another important parameter, $x$, 
which describes field fluctuation. 
In papers of Jokipii (1987) \cite{jokipii} and 
Ellison et al. (1995) \cite{ellison95}, 
they use $\xi=\lambda/r_{g}$, which simply relates 
to $x$ as $x^{2}=\xi^{2}+1$ (In their original papers, 
they use the notation $\eta$ instead of $\xi$ but 
here we use $\xi$ in order to distinguish it from our 
$\eta(\equiv\cos\alpha_{1})$).  $\xi \sim 1$ 
corresponds to the Bohm diffusion (strong scattering) limit 
and $\xi\rightarrow \infty$ when 
$\kappa_{\bot}\ll\kappa_{\|}$.
According to ref. \cite{ellison95}, the $x$ value is limited by
$x \lesssim v/U_1$, where $v$ is the particle velocity, and in 
quasi-perpendicular shocks $x \lesssim \tan \alpha_1$. 
Our choices $x = 30$ and $\eta_{\rm min} = 1/30 $, which 
will be adopted later, however, do not satisfy the latter 
inequality. The fixed $x$ value is also used for any $\eta$ 
between $\eta_{{\rm min}}$ and 1.

\section{Our model}
Based on the result that the angular dependence of the 
maximum energy is quite significant, we propose a 
simple model. The following three issues are examined: (1)
the chance probability of $\eta$ occurrence, (2) the dependence 
of the injection efficiency on $\eta$ and (3) the $\eta$ dependence 
of the spectral index.
\subsection{Chance probability of $\eta$}
The overall morphology of SNR-shocks is considered to be 
almost spherically symmetric even if the shape of each is 
somewhat distorted. The ambient 
disordered magnetic field lines 
over a wide range intersect the shock normal at various 
angles. 
Moreover, as we described in the section II. B,
there exist SNRs with various ages whose 
magnetic fields cross the shock fronts with a wide variety
of angles. Since both young and old SNRs are considered as sources of 
energetic cosmic rays, averaging
over the various 
ages of SNRs and different shapes of their shock fronts, we assume  
a uniform distribution of $\eta$ in our model. 

The value of $\eta$ is restricted within the range between 
$\eta_{{\rm min}}$ and 1 to satisfy the condition that the 
shock front should not move at the light velocity 
or above in the H-T frame. 
The probability in the width $d\eta$ is then
\begin{equation}
f(\eta) d\eta=\frac{d\eta}{1-\eta_{{\rm min}}}~~~~~
\left(\eta_{{\rm min}}\leq\eta\leq 1\right),
\label{eqn:chance}
\end{equation}
where $\eta_{{\rm min}}=U_{1}/c \sim 1/30$ and 
$(1-\eta_{{\rm min}})^{-1}$ is the normalization factor. 
The range where $0\leq\eta<\eta_{{\rm min}}$ is neglected 
because it occupies only $1/30$ of the whole $\eta$ region. 
Let $\mathcal{E}$ be the energy of a particle with 
$E_{{\rm crit}}\le \mathcal{E} \le E_{\rm cut}$.  
$E_{{\rm crit}}$ and $E_{\rm cut}$ are the maximum energies in 
parallel ($\eta = 1$) and extremely oblique 
(quasiperpendicular: $\eta=\eta_{{\rm min}}$)
shocks, respectively.
A particular value of  $\eta$ corresponds to an unique $\mathcal{E}$ 
since $E_{{\rm max}}(\eta)$ is a monotonic decreasing function 
of $\eta$ as shown in Fig. \ref{fig:emax}. Only oblique 
shocks in the range from the $\eta$ to $\eta_{{\rm min}}$ 
can then serve to give $\mathcal{E}$. The fluxes at $\mathcal{E}$ 
should therefore 
be reduced by the factor $(\eta-\eta_{{\rm min}})/
(1-\eta_{{\rm min}})$.

\subsection{Injection efficiency}
Baring, Ellison and Jones \cite{baring93,baring95} developed  
test particle 
Monte-Carlo simulations with large angle scattering processes
(abbreviated as LA). They showed that the efficiency for injecting
thermal particles into an acceleration mechanism strongly depends 
on obliquity and varies inversely with ${\alpha}_1$ 
(${\Theta}_{\rm Bn1}$ in their notation), in the case of high
Mach numbers. They also checked the $\xi$ ( $\approx$ our $x$)
dependence of the injection efficiency, $\epsilon(\eta)$.
When their results are applied to the present case $x \approx 30$,
the injection efficiency relative to $\eta =1$  
is nearly equal to or less than 0.01 at $\eta =0.6$. Namely
oblique shocks with $\eta$ from $\eta_{\rm min}$ to 0.6 
contribute very little to the acceleration of particles.

Naito and Takahara \cite{naitop,naitom} also attempted 
test particle simulations in oblique shocks. Instead of 
directly presenting $\epsilon(\eta)$, they showed the 
pitch angle (${\mu}$) distribution in both cases of 
LA and pitch-angle scattering (abbreviated as PA) in 
their figures 3 and 4 \cite{naitom}, respectively, for 
$\eta=1.0, 0.75, 0.5, 0.25$. Using the number of 
injected particles for these $\eta$ values, we can estimate 
$\epsilon(\eta)$ by integrating with respect to ${\mu}$ 
from $-1$ to 1 in the downstream, where the ${\mu}$ 
distribution of their simulation agrees well with the 
expected one \cite{naitop}. The relative injection 
efficiency $\epsilon(\eta)$ changes little even if the number 
of accelerated particles in the upstream 
are added, because the numbers are less than those in the 
downstream and also the ratios to $\eta=1.0$ are similar to
those in the upstream. 

For LA, $\epsilon(\eta) = 1,\  0.73,\ 0.49,\ 0.26$, whereas for PA
$\epsilon(\eta) = 1,\ 0.81, \ 0.59, \ 0.31$  are evaluated corresponding 
to $\eta = 1, \ 0.75, \ 0.5,  \ 0.25$, respectively. 
It is argued \cite{ellison90} that LA is adequate for 
$\frac{{\delta}B}{B} \sim 1$,
while PA is acceptable for $\frac{{\delta}B}{B}<0.1$. In our case,
$(\frac{{\delta}B}{B}) ^2 \sim \frac{1}{x} = \frac{1}{30}$ is somehow
between LA and PA, and $\epsilon(\eta)$ is similar in both cases. 
Therefore, we may simply take,
\begin{equation}
\epsilon(\eta) = \eta  
\label{eqn:inject}  
\end{equation}

The detailed reason why the above two works present 
very different results on $\epsilon(\eta)$, is not 
clear. The input for simulations,
however, differs for the two approaches. In Ellison et al. 
\cite{ellison95,ellison96}, 
injected particles are thermal, the energies of accelerated
particles are up to about 10 MeV, and $U_1$ is slow, say
$5\times 10^5$ m/s. Ellison has also presented 
the application to the real SNR \cite{ellison01}. He considered 
somewhat older objects which are in Sedov phase. 
On the other hand, in
Naito and Takahara \cite{naitom}, the Lorentz factors of injected 
particles are fixed at 2, the factors of accelerated particles 
are followed up to $10^{5}$, and  the shock is relativistic: 
$U_1\sim 0.1c$.
In the light of the present treatments or parameters, such as  
$U_1 = 10^7 \ {\rm m}/{\rm s}$  and considering that the energies concerned 
are higher than 1 TeV, we assume Eq. (\ref{eqn:inject}) here.
    
\subsection{Change of spectral indicies}
\label{index.subsec}
\vspace{-0.5ex}
\hspace*{5mm}
In this subsection, we consider the dependence of the 
spectral index on obliquity. It has been frequently 
pointed out that one of the advantages of the first order Fermi acceleration 
mechanism is that it gives a power-law energy spectrum. 
Namely one gets for the differential spectrum
$dJ/dE$, if we take terms of 
fluid velocities up to first order into account, 
\begin{equation}
\frac{dJ}{dE}\propto E^{-q} ~~~~~~~~~
\left({\rm with}~~~~q=\frac{r+2}{r-1}\right).
\end{equation}
For a strong shock ($r=4$), the differential spectral 
index is $q=2$. 
This index is expected to be modified into the final index $\gamma$ 
which is observed. 
$\gamma$ is somewhat steeper than $q$ by considering the 
energy-dependent leakage from the galaxy \cite{leaky}. 
We do not go into detailed discussions of this change in 
this paper, even though different modifications are used 
for different nuclear groups in the next subsection. 

This dependence of $q$ values on $\eta$ is examined. This dependence 
is assumed to apply to $\gamma$ in the same form. Although 
nonlinear back reaction effects of 
accelerated particles reduce $q$ values by about 0.5 at higher energies 
even in parallel shocks \cite{blasi}, only the index dependence on $\eta$ 
is considered here. The dependence has been studied in earlier works
\cite{ostrowski91,ostrowski93}. 
Here, we again refer to Naito and Takahara 
\cite{naitop,naitom}. In ref. \cite{naitop}
they showed that $q$ is practically 2 at $\eta =1$, and diminishes 
as $\eta$ decreases except for $\eta \approx \eta_{\rm min} $ being rather 
independent for the three cases $\frac{U_1}{c} = 0.01, 0.03, 0.1$.
It is also shown \cite{naitom}, for $\frac{U_1}{c} = 0.1$ in the case of
LA, $q$ is 2.08, slightly larger than 2, at $\eta =1$, and almost 
linearly decreases as $\eta$ decreases. Then $q$ reaches a minimum
of 1.5 at $\eta \approx 0.15$. For $\eta_{\rm min}< \eta < 0.15$, 
$q$ starts to increase rapidly. While for the PA case, $q$ is 
nearly equal to 
2.1 at $\eta=1$. As $\eta$ decreases, $q$ decreases monotonically 
and reaches 1.0
as $\eta$ approaches $\eta_{\rm min}$. In both cases the energy
spectra become harder as shocks are more oblique.

We choose the LA model except for $\eta \approx \eta_{\rm min}$ and assume
simply that $q$ is proportional to $\eta$ such as,
\begin{equation}
q ({\eta}) = a{\eta} + b             
\qquad\quad
( 1 < {\eta} < \eta_{\rm min}) ,
\label{eqn:q}
\end{equation}
where $a$ = 0.68, $b$ =1.41 and $\eta_{\rm min} = \frac{1}{30}$.
This dependence of power indices on obliquity enhances 
intensities of accelerated particles by oblique shocks.
The present choice leads to an underestimate of this enhancement.
For the energies of particles $\mathcal{E}$ less than $E_{\rm crit}$,  
the enhancement factor should be unity, because of $q$
levelling off by summing up over the whole region of $\eta$, i.e. from 1
to $\eta_{\rm min}$.

For $E_{\rm crit}< \mathcal{E} < E_{\rm cut}$, 
there is a one-to-one correspondence of $\eta$ 
to $\mathcal{E}$ as mentioned above.
Taking into account the injection efficiency given by 
Eq. (\ref{eqn:inject}),
we introduce a correction factor 
\begin{equation}
\zeta(\eta) = \frac{2}{1- {\eta_{\rm min}}^2}\int_{\eta_{\rm min}}
^{\eta}\ {\eta'}\ \left( \frac{\mathcal {E}}{E_{\rm crit}} 
\right)^{-q({\eta'})+q(\eta_{\rm a}) }\ d{\eta'} ,
\label{eqn:zeta}
\end{equation}
where $\eta_{\rm a} = (1+\eta_{\rm min})/2 $ is the average value
of ${\eta}$. The factor $-q({\eta}')+q(\eta_{\rm a})$ leads to
a deviation of indices between $1 \sim \eta_{\rm min}$ and 
$\eta \sim \eta_{\rm min}$ and hence, $b$ in Eq. (\ref{eqn:q}) 
is cancelled.
When $\mathcal{E} = E_{\rm crit}$, 
that is to say $\eta = 1$, $\zeta(\eta = 1) = 1$ due to the 
normalization factor $\frac{2}{1- {\eta_{\rm min}}^2}$. 
The $\eta$ dependence of indices compensates partially the injection 
efficiency but not completely. So $\zeta(\eta)$ is less than unity such as
for $\eta_{\rm min} < \eta < 0.8,\ 0.7 < \zeta(\eta) < 0.8.$

In addition to $\zeta(\eta)$, the chance probability 
of $\eta$ given by Eq. (\ref{eqn:chance}) should be combined. 
Since the probability is independent of $\zeta(\eta)$, after 
integrating Eq. (\ref{eqn:chance}) we 
obtain a final correction factor:
\begin{equation}
g(\eta) = \frac{\eta-\eta_{{\rm min}}}{1-\eta_{{\rm min}}}\ \zeta(\eta)
\label{eqn:correction}
\end{equation}  
In Fig. 3, $\zeta(\eta)$, $\frac{\eta-\eta_{{\rm min}}}
{1-\eta_{{\rm min}}}$ and $g(\eta)$ are sketched as 
dashed, dotted and solid curves, respectively.  
 
In summary, the differential energy spectra can be 
expressed as:  
\begin{eqnarray}
\frac{dJ}{dE} &=& CE^{-\gamma}~~~~~~~~~~~~~~~~~~~~   
(E< E_{{\rm crit}})\nonumber\\
 &=& CE^{-\gamma}~g(\eta)~~~~~~~
(E_{{\rm crit}}\leq E\leq E_{{\rm cut}})
\nonumber\\
 &=&0~~~~~~~~~~~~~~~~~~~~~~~~~~(E_{{\rm cut}}<E),
\label{eqn:flux}
\end{eqnarray}
where $C$ and $\gamma$ are fixed according to six 
nuclear groups in the next section. $E_{{\rm crit}}$ 
is given by Eq. (\ref{eqn:rigidity}) and 
$E_{{\rm cut}}=1.84Z\times 10^{17}$ eV.  
From Eq. (\ref{eqn:flux}), one 
can see that the curve of the flux is differentiable 
changing from 
linear to a smooth bent above $E_{{\rm crit}}$ in a 
log-log diagram. 
 
The velocity of the shock front becomes superluminal 
in the H-T frame in the region $0\leq\eta <\eta_{{\rm min}}$, 
which we have neglected. 
If particles are accelerated effectively by some other special 
mechanism, however, we should take this range into consideration. 

\section{Results}
We calculate differential fluxes of various mass groups. 
The obtained fluxes are multiplied by $E^{2.5}$ in order 
to clarify the change of the spectral indices. The multiplied 
flux for each chemical component is normalized to the 
experimental data at $E=1$ TeV, where $E$ is the total 
(mass plus kinetic) energy per particle, to give 
\begin{equation}
\Psi=\left[E^{2.5}\frac{dJ}{dE}\right]_
{E=1{\rm TeV}}=\phi E^{-\beta},
\label{eqn:abflux}
\end{equation}
where $\beta=\gamma -2.5$. 
Here we take as an assumption two different types of 
values of $\Psi$ and $\beta$ as shown in Table \ref{tbl:condition}. 
Parameters with subscript ``HEGRA'' have been taken from 
the analysis of the
HEGRA air shower experiment \cite{hegra} and slightly 
modified such that the heavy and very heavy nuclei (HVH) 
parts are separated into $\Psi$ of sub-Fe and Fe with 
the ratio $0.125$ to $1.0$ and with the same $\beta$.  
The values with suffix ``Ours'' have been determined so 
as to fit well various experimental data around $E=1$ TeV. 

We do not discuss in detail why some different values of 
$\Psi$ and $\beta$ are taken for groups of elements as 
in Table \ref{tbl:condition}. The most notable difference 
of these two models is the exponent $\beta$ of the Fe spectrum. The 
resultant all-particle spectrum with ``HEGRA'' parameters 
shows a more rapid decrease for high energies.  
Since $x \sim (B/\delta B)^2 $ which is estimated to be 
between 10 and 100, for simplicity, we fix hereafter the 
value of $x$  
at about its geometrical mean of $x=30$ in the calculation 
of Eq. (\ref{eqn:emax1}). 

\subsection{The energy spectra of primary cosmic ray components}
In Fig. \ref{fig:spphe} we plot the calculated spectra of 
p and He against the kinetic energy $E_{k}$ per nucleon. The 
theoretical spectra with ``Ours'' parameters are expressed 
by solid curves and those with ``HEGRA'' parameters by dashed 
ones.
The energy spectrum of p starts to bend due to the $g$ factor 
in Eq. (\ref{eqn:flux}) at the energy:  
$E_{k,{\rm crit}}({\rm p})\simeq 1.25\times 10^{5}$ GeV/n. 
This steepening of the spectral index is visible  
in the recent data of the Tibet air shower experiment 
\cite{tibet1,tibet2,tibet3}. Here we cite the data of 
Tibet (2001) labelled proton dominant (PD) model in 
their paper. But their results from heavy dominant 
model are almost the same with those from PD.
For He, $E_{k,{\rm crit}}({\rm He})=Z({\rm He})/A({\rm He})
\times E_{{\rm crit}}({\rm p})-m_{{\rm p}}\simeq 6.3\times 
10^{4}$ GeV/n. 
 JACEE data \cite{jacee} seem to be consistent with this 
$E_{k,{\rm crit}}({\rm He})$,  while RUNJOB's fluxes \cite{runjob} 
are somewhat lower but their highest $E_{k}$ does not reach 
$E_{k,{\rm crit}}({\rm He})$. 

We also present spectra of the CNO, middle, and Fe-group 
in Fig. \ref{fig:splmh}. ``middle'' corresponds 
to the NeMgSi-group and the Fe-group to our ``sub-Fe'' plus 
``Fe'' in Table \ref{tbl:condition}. The two types of 
curves are defined by the same notation as in Fig. \ref{fig:spphe}. 
The fluxes of middle and those of Fe-group have been downscaled by factors 
of 10 and 100 with respect to the original values, respectively. 
The slow changes of spectral indices of these groups 
above $E_{k,{\rm crit}}\simeq 6.3\times 10^{4}$ GeV/n 
for CNO and $6.2\times 10^{4}$ GeV/n for middle 
are not clearly seen from the data, because it is difficult 
to obtain such data by direct observations. 

For Fe, $E_{k,{\rm crit}}\simeq 5.8\times 10^{4}$ GeV/n,  
the JACEE point with the highest energy ($\simeq 10^{5}$ GeV/n) 
shows a large value but with a large error. On the other 
hand, the recent Tibet data around $E_{k}\simeq 10^{6}$ 
GeV/n \cite{tibet4} show that the spectrum is constant 
or has only a small bent in the diagram of Fig. \ref{fig:splmh}. 
This favors Ours choice over HEGRA's. 

We plot the theoretical curves of spectra for various nuclear 
groups with Ours choice of parameters versus the total  
energy per particle in Fig. \ref{fig:spvar}. One can see 
that the spectra of lighter nuclei start to bend at 
lower energies. This is just due to Eqs. (\ref{eqn:emax2}) 
and (\ref{eqn:flux}). Namely, the changes occur at a fixed 
rigidity of $1.25\times 10^{5}$ GV as shown in 
Eq. (\ref{eqn:rigidity}).
Summing the fluxes of all these groups, we obtain the flux 
of all particles which is plotted as ``total'' in this figure. Since the  
spectra of each group are smooth (differentiable) as mentioned 
above, the total spectrum should also be smooth. The knee behavior in 
our model comes from the $\eta$-dependence of $E_{{\rm max}}$ due 
to the oblique shock acceleration expressed in the second 
line of Eq. (\ref{eqn:flux}). 
This is different from the HEGRA analysis \cite{bernlo}, 
Biermann \cite{biermann}, Stanev et al. \cite{stanev}
and Erlykin and Wolfendale \cite{erlykin00}.
In the first paper, the knee is explained by introducing an 
artificial break 
of slope at a fixed rigidity. The authors of the next two papers, 
interpret the knee as the superposition of sources from different 
phases of SNRs. In the last reference, the authors argue that
a sudden steepening at the knee and the explosion of a single, 
recent, nearby supernova is most likely to explain cosmic rays 
around the knee.

In Fig. \ref{fig:sptot} the two curves of the total flux, 
i. e. with Ours (the same as in Fig. \ref{fig:spvar}) and HEGRA 
choices of parameters, are compared with experimental data. 
As seen in this figure, Ours curve fits well 
Tibet \cite{tibet4} and Akeno data \cite{akeno} up to several 
times 
$10^{8}$ GeV, while the curve with HEGRA parameters is in good 
agreement with DICE \cite{dice,kieda} and CASA-MIA \cite{casamia}, 
in which both $E$ and $\Psi$ are about 30$\%$ lower than 
the Akeno data. The KASCADE experiment presented a preliminary 
result with a four component assumption (H, He, C, and Fe) 
\cite{kascade-new}. It is shown that the position of the knee depends 
on rigidity. That is also one of our conclusions.   
Anyway, the total flux can be explained 
at least up to $10^{8}$ GeV by the present model. 
Above that energy, the solid curve falls below the 
AKENO data. Other sources than SNRs in Galaxy may be 
needed. 
    
We express the two curves in Fig. \ref{fig:sptot} in single 
power laws below and above $E_{{\rm knee}}$ by using the least 
square method. For HEGRA and Ours choices 
$\gamma=2.713\pm 0.006$ and $2.664\pm 0.002$, respectively,  
between $1.5\times 10^{5}$ GeV and $1.5\times 10^{6}$ GeV; 
and $\gamma=3.248\pm 0.012$ and $3.092\pm 0.008$, respectively, 
between $6\times 10^{6}$ and $6\times 10^{8}$ GeV are obtained. 
These values depend on the adopted energy range; but they can be  
compared with experimental data. For example, 
Tibet $\gamma$ \cite{amenomori} gives 
$2.60\pm 0.04$ below and $3.00\pm 0.05$ above $E_{{\rm knee}}$ 
and CASA-BLANCA \cite{casablanca} $2.72\pm 0.02$ and $2.95\pm 0.02$.     

\subsection{The energy dependence of chemical composition}
Using the results of energy fluxes, we examine how the 
chemical composition of cosmic ray particles changes 
with energy. In Fig. \ref{fig:rate} the flux ratio of p 
plus He to all particles is plotted against the energy 
per particle. Theoretical curves suggest a change with increasing 
energies from mixed composition to heavy particle dominance.  
The curve with HEGRA parameters lies almost within the 
uncertainties of experiments quoted here. The relative 
abundance of light elements, 
however, tends to increase again beyond the energy of 
several times $10^{6}$ GeV from experiments of DICE and HEGRA. 

As a commonly used indicator of the composition, we take the 
average value of the logarithm of the mass number $A$, as 
\begin{equation}
<\ln A>\equiv\frac{\displaystyle{\sum_i} \:  f_{i}(\ln A_{i})}
{\displaystyle{\sum_i}  f_{i}} , 
\end{equation} 
where $f_{i}$ is the flux of species $i$ which denotes 
each elemental group as shown in Table \ref{tbl:condition}. 
Our two curves of $<\ln A>$  versus energy $E$ are shown and 
compared with various data in Fig. \ref{fig:comp}. Both 
curves monotonically increase with energy, which is naturally 
confirmed by the present model. The HEGRA parametrization gives lower values 
than Ours in the whole energy range. 

In Fig. \ref{fig:rate} experimental data scatter and have 
large uncertainties especially near and above the knee. 
A similar trend 
is seen in Fig. \ref{fig:comp}. This is due to the fact that 
ground-based measurements can study the composition only indirectly. 
The values of $<\ln A>$ significantly depends on the experimental 
observables to be treated  and on the hadronic interaction models 
to be used in simulation processes. Fairly different 
$<\ln A>$ values 
are reported by various groups. Some authors \cite{dice} say 
that $<\ln A>$ decreases in energy regions higher than the 
knee. Other authors \cite{casablanca} show that $<\ln A>$ has even 
a dip near 
$E_{{\rm knee}}$. Most of the other data indicate that $<\ln A>$
increases with $E$. 
Preliminary results of the KASCADE experiment 
\cite{kascade-new} show a sharp increase in logarithmic mass 
above several times of $10^{6}$ GeV. On the other hand, the data 
extracted from equi-intensity curves depicted by 
Chacaltaya II \cite{chacaltaya2} are somewhat closer to Ours. 
In the present stage we cannot reach 
a definite conclusion about whether our curves are in good agreement 
with data or not. 

\section{Concluding remarks}
We consider the acceleration caused by oblique shocks in which outer 
magnetic field lines cross the shock normal at any angle $\alpha_{1}$. 
We calculate $E_{{\rm max}}$ as a function of $\eta=\cos\alpha_{1}$, 
where $E_{{\rm max}}$ is the maximum energy that particles can achieve 
(see Eq. (\ref{eqn:emax2})). It is shown that $E_{{\rm max}}$ with 
extreme obliquity is more than 1400 times compared to that with 
parallel shocks 
($\eta=1$). In the present model, $\eta$ is distributed 
uniformly and the $\eta$ dependence of the injection efficiency and
the change of spectral indices are taken in to account.
We choose the two parameter sets of power indices and absolute 
fluxes at $10^{12}$ eV for various elemental groups shown in 
Table \ref{tbl:condition}. Then the spectra and compositions 
of cosmic rays are calculated up to much higher energies, 
say $\sim 10^{17}$ eV.  
It is shown that 
$E_{{\rm crit}}=1.25Z\times 10^{14}$ eV, where spectral 
curves start to bend over and continue up to $E_{{\rm cut}}
\simeq 1470 \times E_{\rm crit}$. These results 
fit the proton spectrum well and explain a smooth knee 
behavior around $3\times 10^{14}$ eV (see Fig. \ref{fig:spphe}). 

As to the total flux, both curves in Fig. \ref{fig:sptot} exhibit a 
smooth knee behavior. The ``HEGRA'' choice fits well DICE and 
CASA-MIA data, while ``Ours'' choice reproduces well Tibet 
and Akeno data up to several times $10^{17}$ eV. However, since 
our model is limited by $E_{{\rm max}}$, for energies higher 
than $\sim 10^{17}$ eV, more energetic sources of cosmic rays  
should be necessary. Favorite candidates for the sources of 
cosmic rays far above the knee may be 
microquasers, pulsars, active galactic nuclei \cite{protheroe}, 
gamma-ray bursts (GRBs) \cite{dermer} or other hitherto unknown sources.

As for the composition, our value of $<\ln A>$ increases with the 
energy per particle as shown in Fig. \ref{fig:comp}. 
Data from various experiments scatter and have large uncertainties, 
particularly above $E_{{\rm knee}}$. Thus the situation 
remains uncertain. 
Recent results by Fly's Eye, Haverah Park, and HiRes groups 
provide some interesting data of composition in the high energy 
region above $10^{8}$ GeV as summarized in ref. \cite{horandel}. 
Data of Haverah 
Park and HiRes show that light components are abundant 
towards $10^{9}$ GeV, while a relatively heavy composition 
is suggested by Fly's Eye. These may also suggest some 
other origins apart from SNRs. 

In the present paper, we fixed $x$, $B_{1}$, $R_{sh}$, 
$r$ and $U_{1}$ (i. e. $\eta_{{\rm min}}$) in Eq. 
(\ref{eqn:emax2}) to evaluate the dependence of $E_{{\rm max}}$ 
on $\eta$ quantitatively and to obtain numerical values 
which could be directly compared with data. Generally, 
there are various shocks at SNRs where these quantities 
differ considerably depending on their ages, sizes and so on. 
However, as far as 
we are concerned with the whole or averaged behavior of 
cosmic rays which are observed, our choices and fixing 
of these quantities seem to be appropriate.  
We neglect the extremely quasi-perpendicular case in which the 
de-Hoffmann-Teller frame cannot be used. 
These are problems remaining to be solved. 

\begin{acknowledgments}
We are indebted to T. Shibata at Aoyama Gakuin 
University for informing us of RUNJOB's data. 
We thank S. Ogio at Tokyo Institute of Technology 
for presenting the new Chacaltaya data. 
We are also grateful to T. Naito at Yamanashi 
Gakuin University for his many useful comments. 
We would like to thank C. Grupen at Siegen University 
for carefully reading the manuscript and correcting 
many mistakes. 
\end{acknowledgments}

\clearpage

\begin{figure}[ht]
\begin{center}
\rotatebox{270}{\scalebox{0.5}[0.5]{\includegraphics{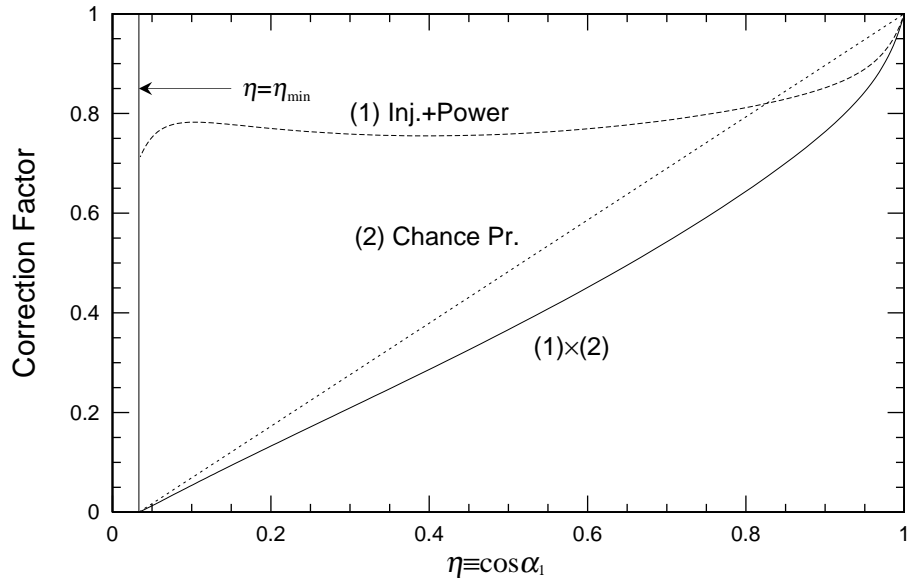}}}
\caption{Correction factors due to the injection 
efficiency plus the change of index power (dashed), 
chance probability (dotted) vs. $\eta$. The final 
reduction factor is expressed by the solid curve.}
\label{fig:cor}
\end{center}
\end{figure}

\begin{figure}[ht]
\begin{center}
\rotatebox{270}{\scalebox{0.5}[0.5]{\includegraphics{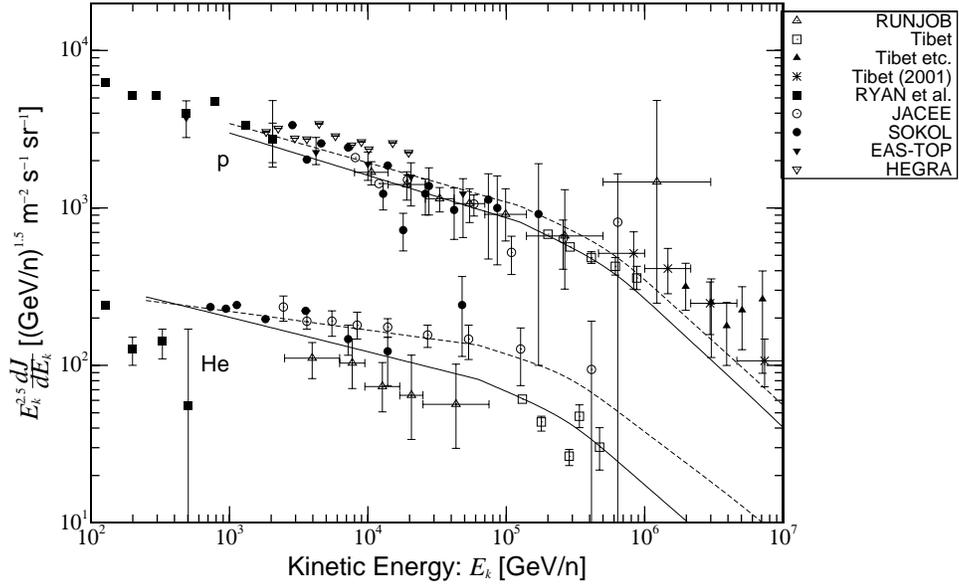}}}
\caption{Comparison of calculated fluxes of p and He 
with experimental data. The horizontal axis represents kinetic 
energy per nucleon (GeV/n). The solid and dashed curves 
represent Ours and HEGRA choices of parameters, respectively 
as shown in Table \ref{tbl:condition}. 
The original data are RUNJOB \cite{runjob}, Tibet etc. 
\cite{tibet1}, Tibet \cite{tibet2}, 
Tibet (2001) \cite{tibet3}, 
EAS-Top \cite{eas-top}, and for others 
see ref. \cite{runjob}. }
\label{fig:spphe}
\end{center}
\end{figure}

\begin{figure}[ht]
\begin{center}
\rotatebox{270}{\scalebox{0.5}[0.5]{\includegraphics{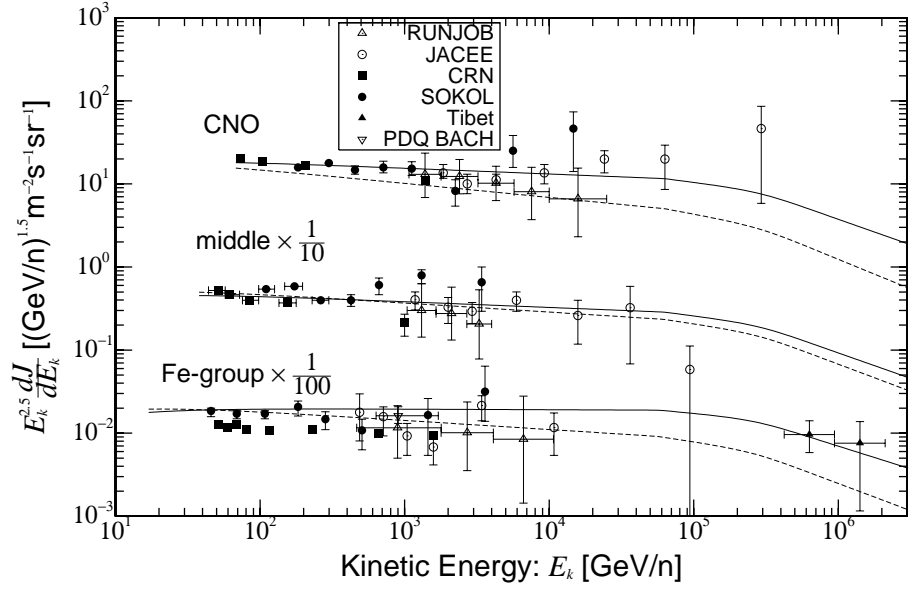}}}
\caption{Comparison of calculated fluxes of CNO-, middle, 
and Fe-groups with experimental data. The original fluxes 
of middle and of Fe-group are multiplied by $1/10$ 
and $1/100$, respectively. The solid and dashed curves 
represent the same notation as in Fig. \ref{fig:spphe}. Experimental data 
are taken from RUNJOB's compilation \cite{runjob}, 
Tibet \cite{tibet4} and PDQ BACH \cite{clem}.}
\label{fig:splmh}
\end{center}
\end{figure}

\begin{figure}[ht]
\begin{center}
\rotatebox{270}{\scalebox{0.5}[0.5]{\includegraphics{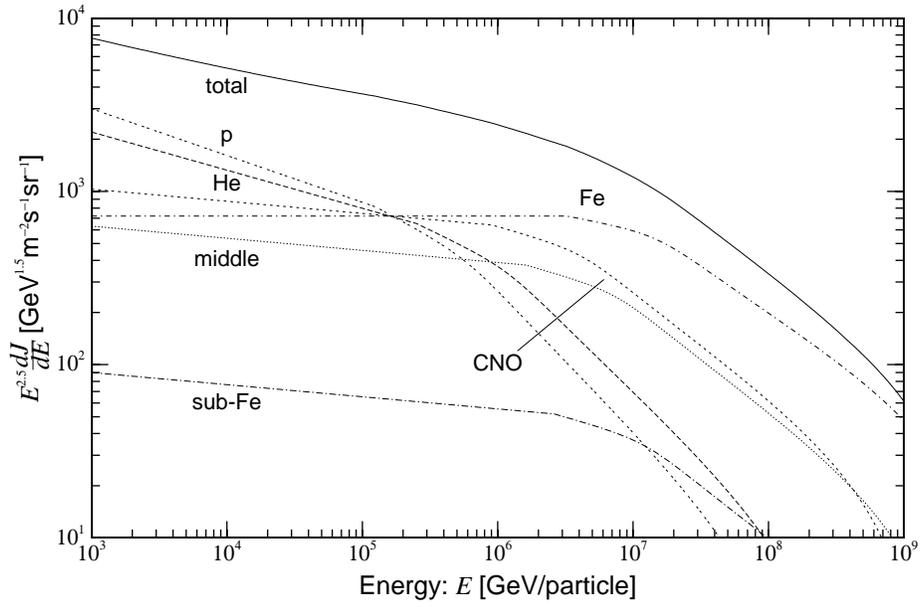}}}
\caption{The predicted fluxes of total, proton, He, and other 
nuclear groups in the case of ``Ours'' parameter choices shown 
in Table \ref{tbl:condition}.}
\label{fig:spvar}
\end{center}
\end{figure}

\begin{figure}[ht]
\begin{center}
\rotatebox{270}{\scalebox{0.5}[0.5]{\includegraphics{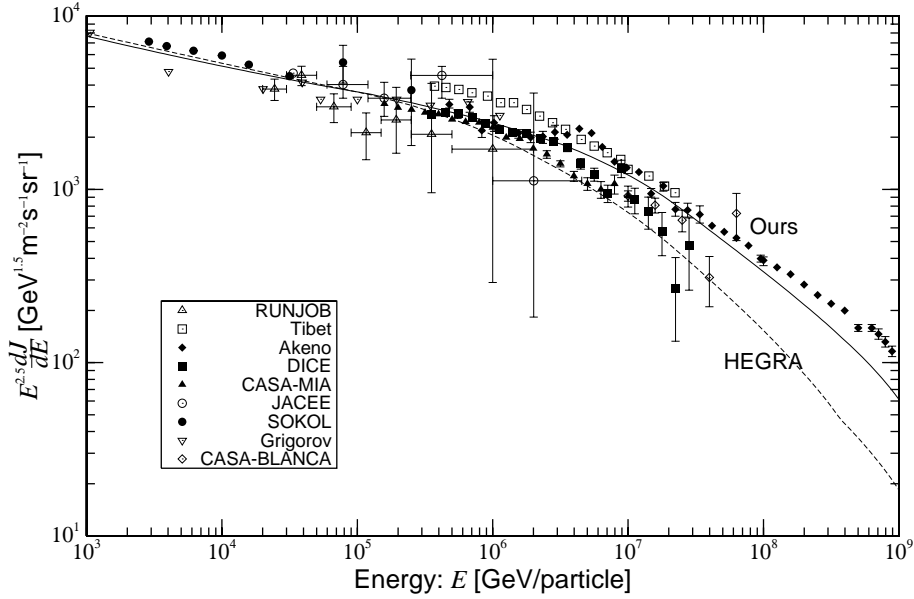}}}
\caption{Comparison of all-particle spectrum versus total 
energy per particle between experimental data from various 
groups and the present calculation. The solid and dashed 
curves represent the same as in Fig. \ref{fig:spphe}. The 
data are Tibet \cite{tibet1}, AKENO \cite
{nagano, akeno}, DICE \cite{dice, kieda}, 
CASA-MIA \cite{casamia} and CASA-BLANCA 
\cite{casablanca} (here only five data points with higher 
energies are cited in order to avoid overcrowding since 
other 16 points of lower energies mostly overlap with 
CASA-MIA data). Others are cited from RUNJOB's compilation 
\cite{runjob}.}
\label{fig:sptot}
\end{center}
\end{figure}

\begin{figure}[ht]
\begin{center}
\rotatebox{270}{\scalebox{0.5}[0.5]{\includegraphics{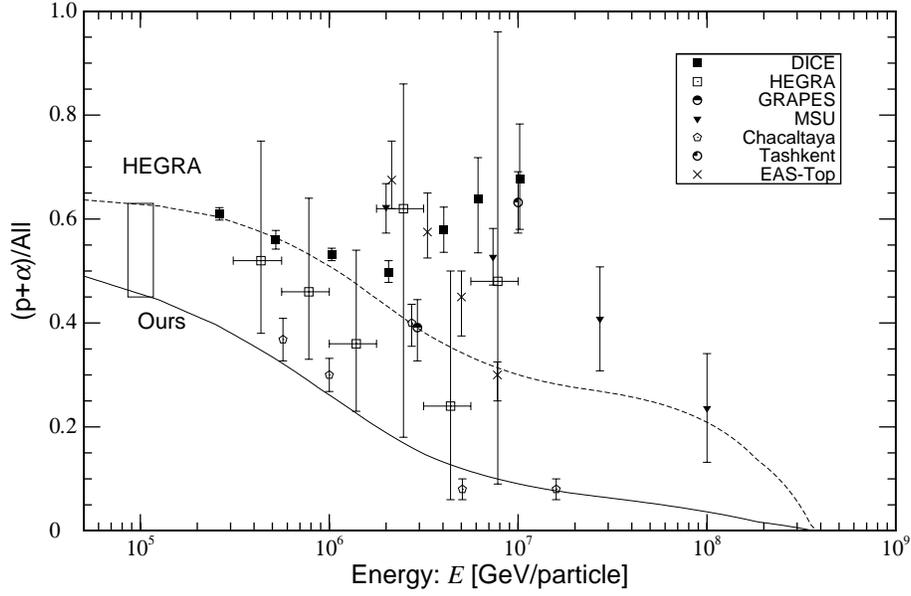}}}
\caption{Energy dependence of $({\rm p}+\alpha)/{\rm All}$. Two 
predicted curves which come from two choices of parameters 
are the same as in Fig. \ref{fig:spphe}. Experimental data are as 
follows: HEGRA \cite{hegra3}, DICE \cite{dice}, 
EAS-Top \cite{eas-top2}, and others 
are Watson's compilation \cite{watson}. The large rectangle area 
shows direct measurement.}
\label{fig:rate}
\end{center}
\end{figure}

\begin{figure}[ht]
\begin{center}
\rotatebox{270}{\scalebox{0.5}[0.5]{\includegraphics{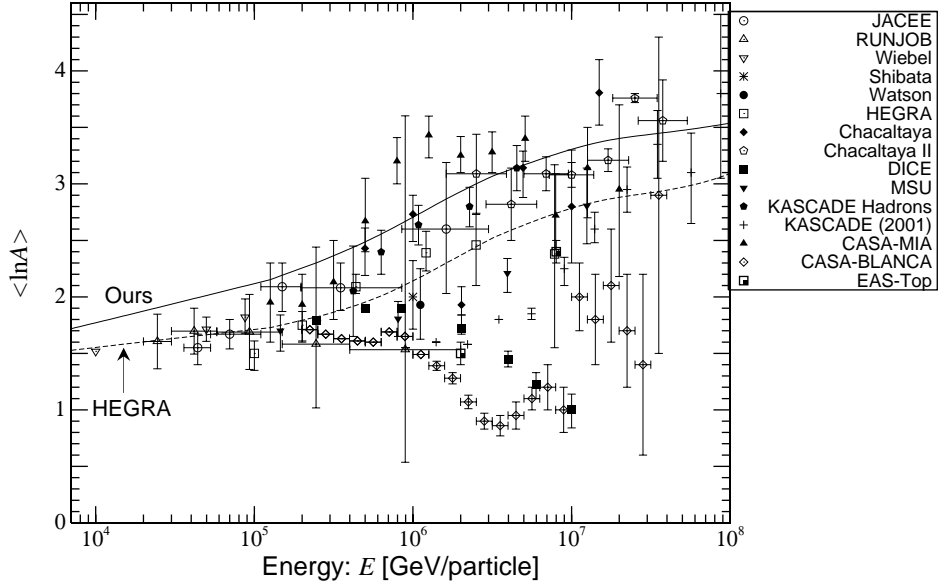}}}
\caption{Energy dependence of chemical composition of 
primary cosmic rays in term of $<\ln A>$. The solid 
and dashed curves come from Ours and HEGRA choices of 
parameters, respectively as shown in 
Table \ref{tbl:condition}.
Experimental data are as follows: JACEE \cite{jacee},  
RUNJOB \cite{runjob}, DICE \cite{dice} ($X_{{\rm max}}+$Muon), 
CASA-MIA \cite{casamia2}, 
CASA-BLANCA (by QGSJET interaction model) \cite{casablanca}, 
EAS-TOP \cite{eas-top2}, 
KASCADE hadrons\cite{kascade}, KASCADE (2001) (preliminary data) 
\cite{kascade-new}, 
and Chacaltaya and Chacaltaya II are cited from 
\cite{chacaltaya} and \cite{chacaltaya2}, respectively.
About half of recent  data from KASCADE (2001) and 
Chacaltaya II are plotted to avoid confusion.} 
\label{fig:comp}
\end{center}
\end{figure}
\clearpage

\begin{table}[h]
\caption{Two choices of parameters for calculation at 1 TeV 
for the absolute value $\Psi$ and index $\beta$ in the form of Eq. 
(\ref{eqn:abflux}). $\Psi$ is in  
[GeV$^{1.5}$m$^{-2}$s$^{-1}$sr$^{-1}$].}
\label{tbl:condition}
\vspace{1em}
 \begin{tabular}
 {|l|c|c|c|c|c|c|} \hline
  \makebox[20mm]{Elements} &
  \makebox[15mm]{$Z$} &
  \makebox[15mm]{$A$} &
  \makebox[20mm]{$\Psi_{{\rm HEGRA}}$} &
  \makebox[20mm]{$\beta_{{\rm HEGRA}}$} &
  \makebox[20mm]{$\Psi_{{\rm Ours}}$} &
  \makebox[20mm]{$\beta_{{\rm Ours}}$} \\ \hline
 \hline
 p & 1.0 & 1.0 & 3447 & 0.25 & 3000 & 0.27 \\ \hline
 He & 2.0 & 4.0 & 2087 & 0.12 & 2200 & 0.22 \\ \hline
 CNO & 7.26 & 14.5 & 885 & 0.17 & 1030 & 0.07 \\ \hline
 Middle & 12.8 & 25.8 & 692 & 0.11 & 630 & 0.07 \\ \hline
 sub-Fe & 21.0 & 45.0 & 99 & 0.11 & 90 & 0.07 \\ \hline
 Fe & 26.0 & 55.9 & 790 & 0.11 & 720 & 0.0 \\ \hline
 total & & & 8000 & & 7670 & \\ \hline
 \end{tabular}
\end{table}

\newpage 

\end{document}